# Cumulative culture spontaneously emerges in social navigators with imprecise memory

Edwin S. Dalmaijer [1]

**Affiliation**

[1] School of Psychological Science, University of Bristol, United Kingdom

**Contact details**

Dr Edwin Dalmaijer, University of Bristol, School of Psychological Science, 12a Priory Road, Bristol, BS8 1TU, United Kingdom. Email: edwin.dalmaijer@bristol.ac.uk

**Note before reading**

This manuscript was shared to invite (preferably constructive) feedback. I welcome any comments, and in particular those that can improve scholarship. This project was sparked by curiosity, and is not in my main area of research. Hence, I am not as familiar with the literature as you perhaps are, and there is a real chance that I missed important papers in my literature searches. Please do let me know if you feel that I missed important work, by others or by yourself. Your suggestions will be used to revise this manuscript before it is submitted to a journal for (further) peer review.

**Acknowledgements**

Thanks to Dr Paul E. Smaldino, Dr Takao Sasaki, and members of the Cultural Evolution Discord for feedback on an earlier version of this manuscript.

**Conflicts of interest**

The author declares that they have no competing interests (financial or other) that could have influenced or appeared to influence the work reported here.

## Abstract

Cumulative cultural evolution occurs when adaptive innovations are passed down to consecutive generations through social learning. This process has shaped human technological innovation, but also occurs in non-human species. While it is traditionally argued that cumulative culture relies on high-fidelity social transmission and advanced cognitive skills, here I show that a much simpler system suffices. Cumulative culture spontaneously emerged in artificial agents who navigate with a minimal cognitive architecture of goal-direction, social proximity, and route memory. Within each generation, naive individuals benefitted from being paired with experienced navigators because they could follow previously established routes. Crucially, experienced navigators also benefitted from the presence of naive individuals through *regression to the goal*. As experienced agents followed their memorised path, their naive counterparts (unhindered by route memory) were more likely to err towards than away from the goal, and thus biased the pair in that direction. This improved route efficiency within each generation. In control experiments, cumulative culture was attenuated when agents' social proximity or route memory were lesioned, whereas eliminating goal-direction only reduced efficiency. These results demonstrate that cumulative cultural evolution occurs even in the absence of sophisticated communication or thought. One interpretation of this finding is that current definitions are too loose, and should be narrowed. An alternative conclusion is that rudimentary cumulative culture is an emergent property of systems that seek social proximity and have an imprecise memory capacity, providing a flexible complement to traditional evolutionary mechanisms.

**Keywords**

## Author Summary

Cumulative culture allows humans to inherit innovations from our ancestors, improve them, and pass it on to next generation through social learning. While initially thought to be uniquely human, cumulative culture is increasingly found in other animals; for example in tool-use by chimpanzees and crows, song-learning by finches and whales; and route-navigation in pigeons. Some argue that high-fidelity communication of information is key to building cumulative culture, and that non-human animals cannot do this. Using artificial agents with minimal cognitive capacities, I show that simple social proximity is enough to communicate memorised routes between experienced and naive individuals, and for naive individuals to improve route efficiency due to regression to the goal. The result is cumulative culture without complex communication.

## Introduction

Cumulative cultural evolution occurs when individuals pass down adaptive innovations through social means (e.g. teaching or copying), leading to progressive increases in fitness over generations (Boyd & Richerson, 1996; Tomasello, 1999). Such socially-transmitted incremental improvements are vital to human technological advancement (Derex, 2022). However, innovations that meet core criteria for cumulative culture also occur in various other species (Mesoudi & Thornton, 2018). It remains unclear whether these species are cognitively advanced, or if cumulative culture is an emergent property in social beings.

For example, homing pigeons (*Columba livia*) are suboptimal navigators that develop and remember idiosyncratic routes when flying alone or in pairs (Biro et al., 2004). While paired birds fly more efficient routes than individuals (Pettit et al., 2013), even greater performance is achieved when pairs fly in *generations* between which experienced pigeons are swapped for a naive ones (Sasaki & Biro, 2017). It could be that pigeons pool information between individuals to innovate, learn and decide through collective intelligence, evaluate performance to prune worse innovations (Sasaki & Biro, 2017); and that they develop intra-pair dynamics of communication and leadership (Valentini et al., 2021a). However, not all species are capable of such cognitively demanding behaviour. Could cumulative culture still develop in the absence of high-fidelity social transmission or cognition?

Here, I introduce a minimal cognitive architecture in artificial agents that are bound by only four rules derived from avian navigation (Figure 1). The first is *goal direction*, akin to birds' solar (Kramer, 1952) and magnetic compasses (Keeton, 1974); the second is *social proximity*, based on the tendency of birds to fly together (Biro et al., 2006); and the third is *route memory*, which in pigeons depends on visual landmarks (Lau et al., 2006) and improves over consecutive flights (Mann et al., 2011). The fourth is *continuity*, to avoid implausibly erratic patterns. Crucially, there is no communal decision-making, evaluation of outcomes, or deliberate social communication.

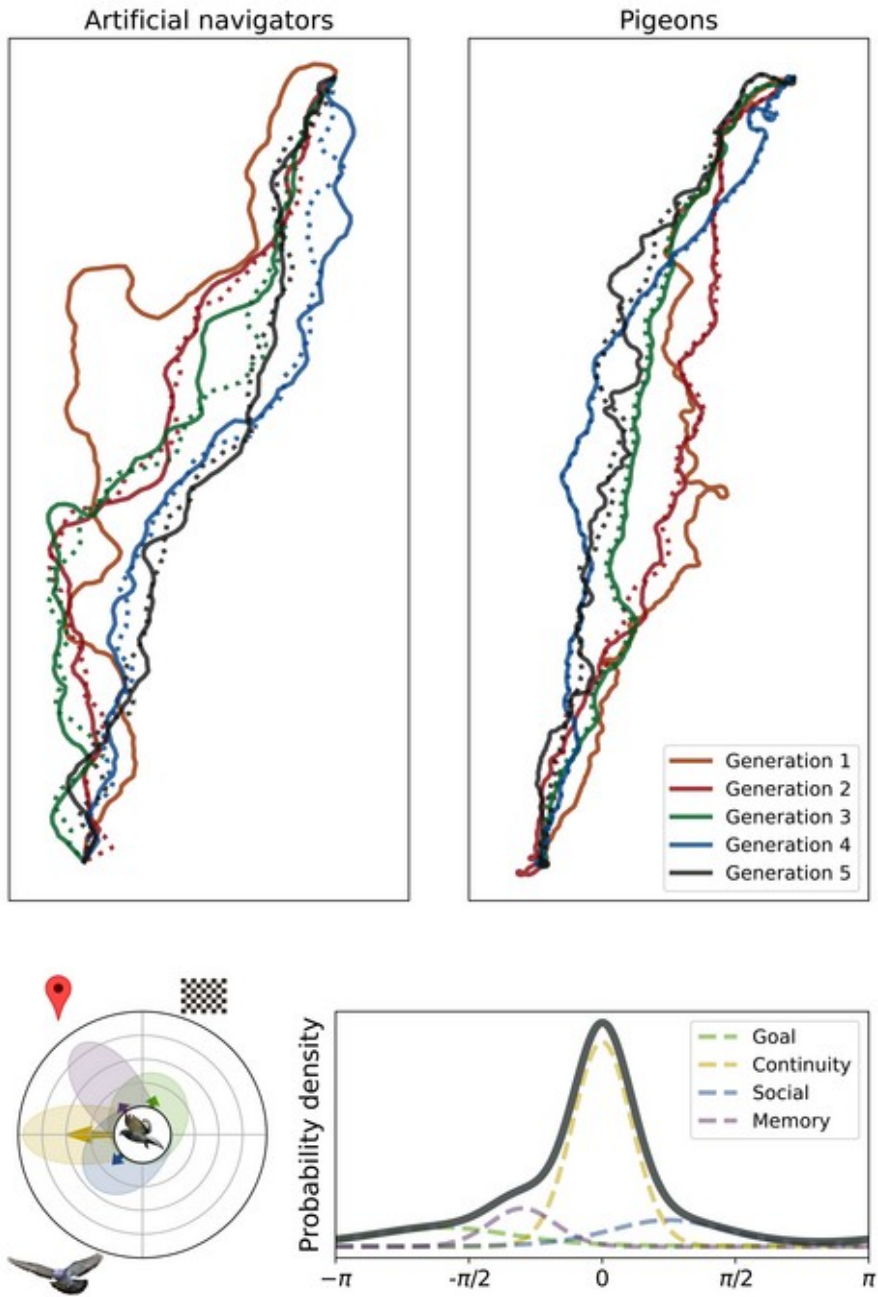

***Figure 1** – The top panel shows paths from artificial agents (introduced here), and from pigeon data published by others (Valentini et al., 2021a). Each line represents the final flight in a generation. The first generation comprises a single individual; a naive individual was added in the second; and in all later generations the most experienced was replaced with a naive individual. Solid lines show lone or experienced individuals, dotted lines show naive ones. The bottom panel shows how agents navigated by sampling from a weighted mixture of Von Mises distributions. These were centred on bearings towards the goal (green), other agents (blue), landmarks along a memorised route (purple), and the previous heading (yellow). Bottom-left shows these distributions in a radial plot, with arrows indicating component centres and weights. Bottom-right shows the distributions and their weighted sum (black).*

The artificial navigator model is a weighted mixture of Von Mises distributions *Φ*, with weights *w* (Equation 1). To produce the next heading in journey *i* at time *t+1*, an agent combines information from time *t* on bearings towards the goal $b_{goal}$, the next memorised landmark $b_{landmark}$, and other agents' estimated future position $\hat{b}_{other}$. As in birds, not all bearings are equally precise, which is reflected in components' spread parameter *κ*. For example, there is larger uncertainty about where the (solar/magnetic compass) goal is compared to where the next (visual) landmark along a well-memorised route is. To prevent unnaturally jerky movements, the final component ensures continuity by sampling from a narrow distribution that is centred on the current heading. For a full account of the algorithm, please refer to *Materials and Methods*.

$$\begin{aligned}h(i,t+1) = &w_{goal}\Phi(b_{goal},\kappa_{goal}) \\ &+ w_{social}\Phi(\hat{b}_{other},\kappa_{social}) \\ &+ w_{memory}\Phi(b_{landmark},\kappa_{memory,i}) \\ &+ w_{continuity}\Phi(h(i,t),\kappa_{continuity})\end{aligned} \quad (1)$$

Agents travelled in three conditions that mapped onto work in pigeons (Sasaki & Biro, 2017): solo, paired, and an experimental condition with generational turnover. In the solo and pair conditions, one or two agents made 60 consecutive journeys. In the experimental condition, a naive replaced an experienced agent every 12 journeys. A total of 10 clean runs were done for each condition, for each set of weight parameters. Spread parameters were fixed at $\kappa_{continuity}$=8.69 (equivalent SD=0.35), $\kappa_{goal}$=1.54 (1.0), $\kappa_{social}$=2.18 (0.80), $\kappa_{memory,1}$=0.85 (2.0) to $\kappa_{memory,5}$=6.78 (0.40), based on model fits for pigeon data.

## Results

Artificial navigators travelled in an "experimental" condition with generational turnover (replacement of an experienced for a naive individual), or in control conditions without turnover (paired or solo). They showed a gradual increase in route efficiency (Figure 2), which was computed as start-goal distance divided by travelled distance (Sasaki & Biro, 2017), and varied between 0 (never reached the goal) and 1 (straight line from start to goal). Parameters could be optimised for final-route efficiency (Figure 2, top) or improvement between generations (Figure 2, centre), and compared well to empirical pigeon data (Figure 2, bottom).

Cumulative culture was quantified as the increase in route efficiency between each generation. This occurred exclusively in the experimental condition (Figures S2 and S3, bottom row; Figure S4, right column), replicating empirical data (Sasaki & Biro, 2017).

Highest final-generation route efficiencies in the experimental condition averaged 0.945, and parameters $w_{goal}$=0.19, $w_{social}$=0.18, and $w_{memory}$=0.31 (pair condition: efficiency=0.941, $w_{goal}$=0.24, $w_{social}$=0.13, and $w_{memory}$=0.32; solo efficiency=0.931, $w_{goal}$=0.27, and $w_{memory}$=0.35). For parameters closest to the final-route efficiency peak ($w_{goal}$=0.20, $w_{social}$=0.175, and $w_{memory}$=0.30), efficiency increased an average of 0.010 per generation. The highest average generational efficiency increase was 0.092, and achieved at $w_{goal}$=0.025, $w_{social}$=0.125, and $w_{memory}$=0.375 in the experimental condition (likely at low $w_{goal}$ because this offers the largest space for improvement).

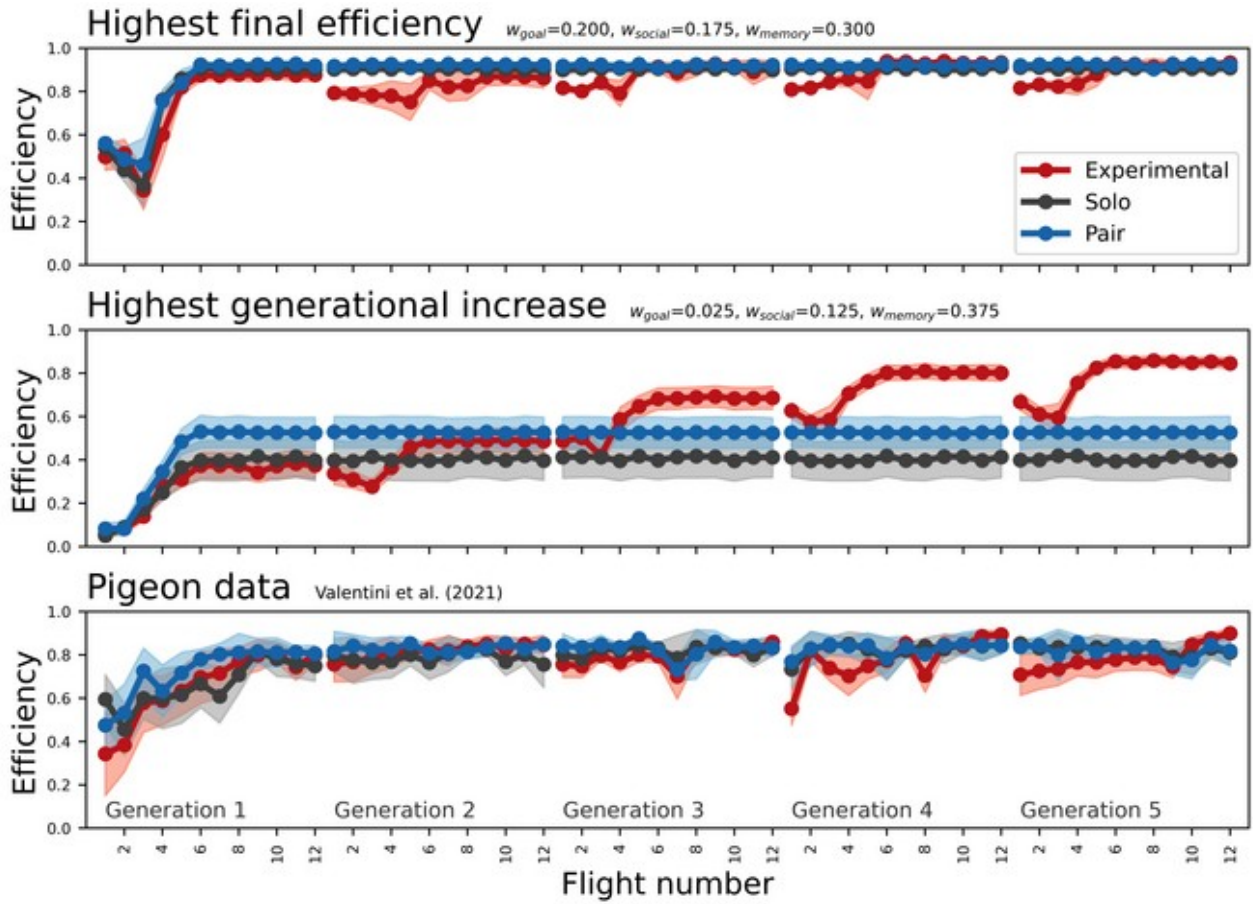

*Figure 2* – *Progression of route efficiency as a function of flight number. The top panel shows results for the optimum for final efficiency, the middle for the optimum for generational improvement, and the bottom panel for pigeon data published by others (8). Lines show mean values over independent runs, with 95% confidence intervals as shaded areas. In the experimental condition, a naive agent replaced an experienced one in each generation; in the solo condition, a single agent made all journeys with no generational turnover; and in the pair condition, two agents journeyed together without turnover.*

**Naive individuals benefit from the experienced**

In the experimental condition, naive individuals benefitted from following an experienced agent with established route memory. Compared to the pair control condition, naive individuals showed more efficient paths (Figure 3). As the relative influence of goal-direction on behaviour ($w_{goal}$) increased, this benefit became limited to early journeys. Efficiency benefits for naive agents improved as the relative influence of social proximity ($w_{social}$) increased, and was highest for intermediate levels of the influence of memory ($w_{memory}$).

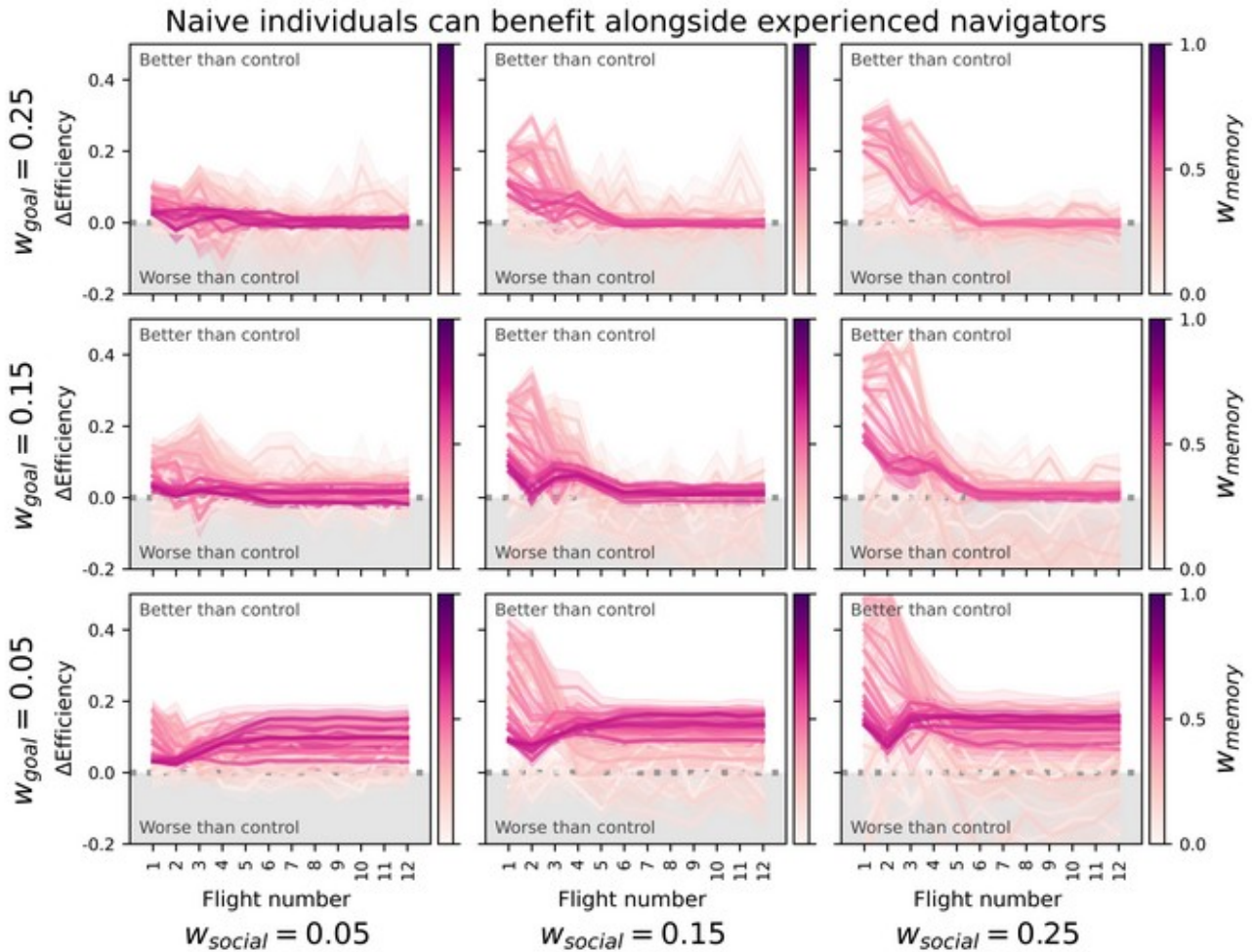

*Figure 3* – *Each panel shows the difference in route efficiency between naive agents in the experimental condition (generational turnover), and the first 12 journeys from agents in the pair control condition (without generational turnover). Positive differences indicate that naive agents had better route efficiency compared to control. Each panel represents a combination of $w_{goal}$ and $w_{social}$ parameters, while darker lines indicate higher levels of $w_{memory}$. Lines represent averages across 10 independent runs, and their shaded areas the 95% confidence interval.*

**Experienced individuals benefit from the naive**

While it is perhaps obvious that naive agents could benefit from following established paths, more surprising was that experienced individuals also benefitted from their naive counterparts. This occurred due to *regression to the goal*. Compared to extreme samples, random samples are more likely to be nearer a distribution's centre; this is regression to the mean. Similarly, experienced agents draw from internal distributions, including for goal-direction and route memory. Naive agents sample from internal distributions too, but do not have route memory yet, and hence are more biased towards the goal than experienced individuals. Because agents aim for social proximity, naive navigators should thus subtly pull experienced agents towards the goal.

    This was born out empirically, as relative bearings for experienced towards naive agents were more likely to also be in the direction of the goal (Figure 4). This was primarily true for lower values of $w_{memory}$, and increased with $w_{social}$. Regression to the goal thus allowed naive agents to memorise slightly more efficient routes than their paired experienced agent.

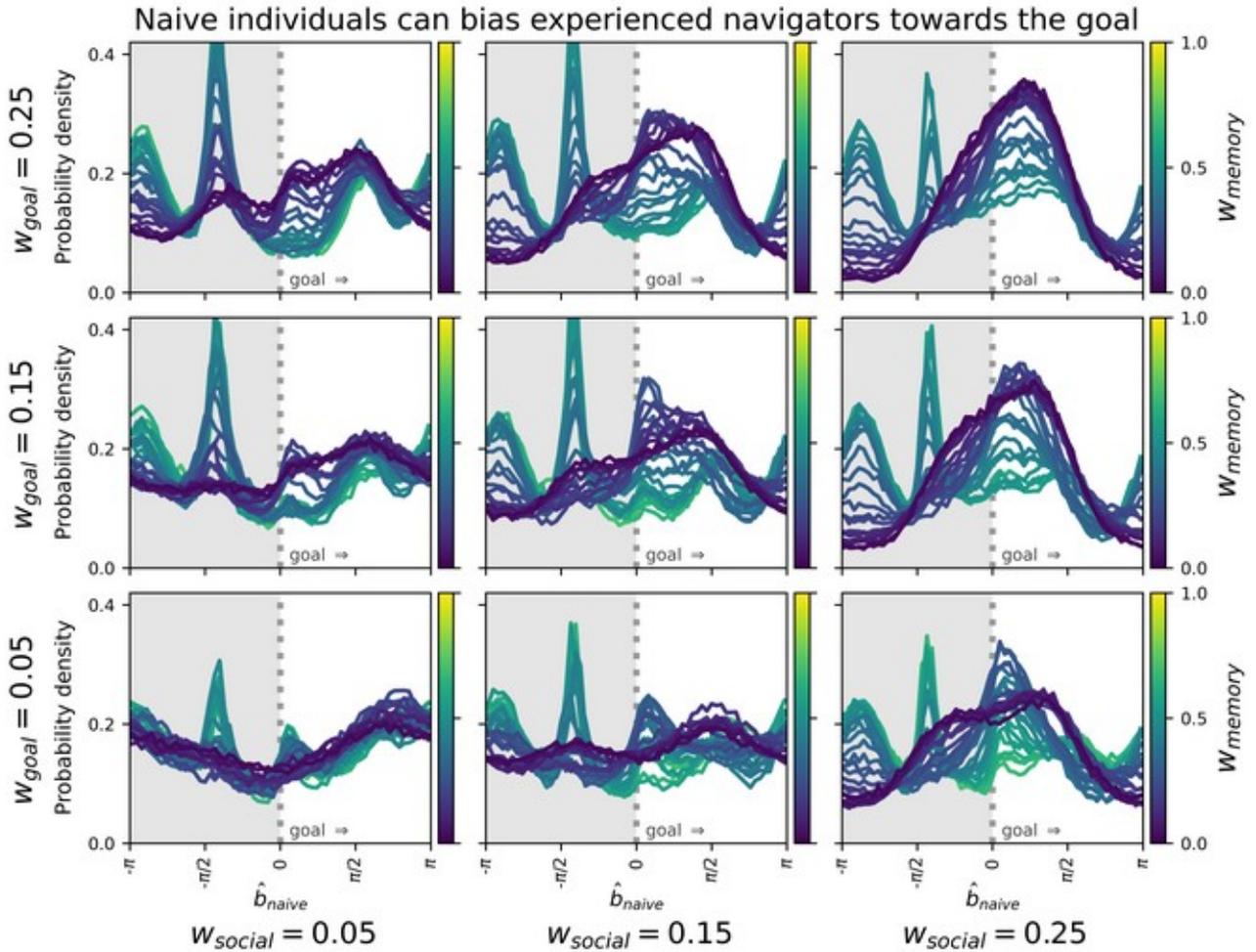

***Figure 4*** *– Each panel shows the distribution of relative bearings towards the naive agent from the perspective of the experienced agent in generations 2-5 of the experimental condition. Positive values on the x-axis indicate bearings towards the goal, and negative values bearings away from the goal. At higher levels of $w_{memory}$ (lighter colours), peaks at -π and -π/2 indicate that naive agents travelled alongside or behind experienced agents. Distributions are generally right-heavy, indicating a bias of naive individuals to be positioned in the general direction of the goal. This tendency increases as a function of $w_{social}$, and to a lesser extent as a function of $w_{goal}$.*

**Control experiments with lesioned agents**

Agents were lesioned in control experiments to investigate which navigation components were necessary for cumulative culture to emerge. Control experiments were identical to those that achieved highest final efficiency or generational efficiency increase in the experimental condition (described above), with the exception that goal-direction, social proximity, or route memory were eliminated. (This was achieved by setting their weight to 0, and compensating by increasing $w_{continuity}$ to ensure weights summed up to 1.)

When goal-direction was lesioned, efficiency was reduced, but inter-generational increases remained. Without goal-direction, agents essentially conducted a random walk search. If they found the goal, this resulted in low-efficiency paths with ample room for inter-generational improvement. Indeed, inter-generational efficiency improvement still occurred in the experimental condition, but not in pair or solo controls. Hence, goal direction was not necessary for cumulative culture.

When social proximity or route memory were lesioned, efficiency was barely reduced, but generational increase was nullified. This suggests social proximity and route memory were crucial for cumulative culture in this model.

*Table 1*

*Efficiency quantifies how close agents were to the direct path from start to goal. Final efficiency is measured in the last generation, and generational increase as the difference between generations. Cumulative culture is reflected by a positive inter-generational increase, and occurs in the "experimental" condition ("pair" and "solo" are control conditions). The "no lesion" column reflects optimal scores; the other columns reflect scores after setting $w_{goal}$, $w_{social}$, or $w_{memory}$ to 0.*

|  | **No lesion** | **Goal lesion** | **Social lesion** | **Memory lesion** |
|---|---|---|---|---|
| **Final efficiency** | | | | |
| experimental | 0.945 | 0.514 | 0.870 | 0.882 |
| pair | 0.941 | 0.218 | 0.888 | 0.881 |
| solo | 0.931 | 0.321 | 0.904 | 0.862 |
| **Generational increase** | | | | |
| experimental | 0.092 | 0.088 | -0.008 | 0.010 |
| pair | -0.001 | -3.27e-5 | 5.49e-5 | 0.007 |
| solo | 4.31e-4 | -1.57e-4 | -8.12e-5 | 0.002 |

**Artificial navigator model fits empirical data**

When fitted to data on pigeons (N=12) flying in stable pairs that was published by others (Valentini et al., 2021a), average parameter estimates were $w_{continuity}$=0.58, $w_{goal}$=0.14, $w_{social}$=0.16, and $w_{memory}$=0.12. That this did not align exactly with agents suggests the artificial navigator model is insufficient to capture the full complexity of pigeon behaviour, which agrees with interpretations put forward by others (Sasaki & Biro, 2017; Valentini et al., 2021a).

## Discussion

The minimal cognitive architecture of goal-direction, social proximity, and long-term memory was sufficient for the emergence of cumulative cultural evolution. It was driven by *regression to the goal* over generations: as agents in a new pair aligned their headings towards each other, experienced agents travelled along a remembered route, while their naive counterparts introduced a subtle goal-directed bias. These artificial navigators met all core criteria for cumulative cultural evolution in an influential framework (Mesoudi & Thornton, 2018): their behaviour **1)** showed variation introduced by interaction between individuals, **2)** was passed on through social interaction (even if implicit), **3)** improved performance, and **4)** repeated over generations.

These results suggest that stepwise improvement between generations can occur even in the absence of high-fidelity communication, when individuals simply seek proximity to others. This is noteworthy, because such ratchet effects have historically been attributed to uniquely human forms of communication (Tennie et al., 2009); if not high-fidelity social learning, then at least strategic social learning and advanced cognitive skills (Zwirner & Thornton, 2015).

The current results suggest that rudimentary ratcheting between generations requires little more than (imprecise) memory and social proximity, arguably one of the lowest-fidelity forms of communication. However, there is a limit to this process: when the direct path between start and goal is reached, no further efficiency improvements can be made. This is optimisation *within* a set of phenomena, i.e. "Type I" cumulative cultural evolution (Derex, 2022).

Artificial navigators did not meet any of the extended criteria for cumulative culture, such as functional dependence, diversification into lineages, recombination across lineages, exaptation, or niche construction (Mesoudi & Thornton, 2018). In fact, their limited cognitive architecture could never achieve the *expansion* of the set of optimised phenomena, "Type II" cumulative cultural evolution (Derex, 2022). While this is described as a *qualitative* difference to distinguish human cumulative culture from that in other animals, simulations suggest that it could arise from a *quantitative* difference in the fidelity of information communication (Lewis & Laland, 2012). While artificial navigators fall below this threshold, avian navigators might not.

In sum, artificial agents with minimal cognition reproduce the intergenerational ratcheting that is core to cumulative cultural evolution. This validates the idea that high-fidelity social transmission is not a requirement for *rudimentary* cumulative culture in animals.

## Materials and Methods

**Artificial navigator model**

Artificial navigators were agents that embarked on journeys from a set starting point to a set goal, although they did not always reach this goal. They were bound by four rules, each implemented as an iterative sampling process from a Von Mises distribution. The centre of each distribution was determined by a bearing, and the spread by certainty of information. At each time point, an agent's heading was updated by sampling each distribution, and computing a weighted circular mean (Equation 1). Weights were set at agent initialisation, and added up to 1. Spread parameters were based on empirical data (see under *"Experimental Design"*).

The first rule was **goal direction**. The centre of this distribution was the bearing towards the goal $b_{goal}$, its spread parameter was $\kappa_{goal}$, and its weight $w_{goal}$. The bearing was computed from the coordinates of the goal ($x_{goal}, y_{goal}$) and agent at time $t$ ($x_t, y_t$) (Equation 2). The purpose of this rule was to orient agents towards the goal.

$$(2) \quad b_{goal} = atan_2(y_{goal} - y_t, x_{goal} - x_t)$$

The second rule was **social proximity**. This distribution's centre was the bearing towards another agent's estimated future position $\hat{b}_{other}$, its spread parameter $\kappa_{social}$, and weight $w_{social}$. This bearing was computed from an agent's position at time $t$, ($x_t, y_t$), and other agent $j$'s expected position at time $t+1$ (Equation 3). The expected position of agent $j$ at time $t+1$ was estimated on the basis of their velocity $v$ (which was kept constant) and their heading $h_{j,t}$ at time $t$ (Equation 4).

$$(3) \quad \hat{b}_{other} = atan_2(\hat{y}_{j,t+1} - y_t, \hat{x}_{j,t+1} - x_t)$$

$$(4) \quad (\hat{x}_{j,t+1}, \hat{y}_{j,t+1}) = (x_{j,t} + v \cos(h_{j,t}), y_{j,t} + v \sin(h_{j,t}))$$

The third rule was **route memory**. This was established during an agent's first journey, in which the positions of 10 landmarks were committed to memory. These landmarks were equally spaced along the travelled route. During consecutive journeys, an agent attempted to fly from one landmark to the next by sampling from a Von Mises distribution with centred on the bearing towards the next landmark $b_{landmark}$, with spread $\kappa_{memory,i}$ for journey $i$, and weight $w_{memory}$ (Equation 5). There were no memorised landmarks in the first journey, so the spread for $\kappa_{memory,1}$ was set to 0, resulting in a completely uniform distribution. For all following journeys, $\kappa_{memory,i}$ was set to 0.27, 0.58, 1.11, 2.18, and then plateaued at 6.78. This was analogous to a linear decrease in standard deviation from 2.0 to 0.4, and was based on model fits to pigeon homing data (see under "*Parameter estimates in pigeons*" in the Results section). Agents proceeded to navigate towards the next landmark *l+1* if they came within a threshold distance of landmark *l*. This threshold was set as 10 times the distance agents could travel between time *t* and time *t+1*.

The gradual improvement in memory precision over several journeys, the anchoring to landmarks, and the number of landmarks per journey were based on Gaussian process models of pigeon navigation (Mann et al., 2011). While the current implementation was less elegant than its inspiration, it was computationally inexpensive, and parsimonious with sampling from distributions of other bearings.

(5) $$b_{memory} = atan_2(y_{landmark,l} - y_t, x_{landmark,l} - x_t)$$

The fourth and final rule was **continuity**. This assured that during journey *i*, an agent's next heading at time *t+1* would be similar to their heading at time *t*. The continuity component was sampled from a Von Mises distribution centred on current heading *h(t)*, with spread parameter $\kappa_{continuity}$, and weight $w_{continuity}$.

Finally, agents set their next heading by drawing random samples *a* from each of the Von Mises distributions described above, and computing their weighted circular mean (Equation 6-8).

(6) $$h(t+1) = \arctan_2(\bar{y}, \bar{x})$$

Where:

(7) $$\bar{y} = \sin(a_{goal})w_{goal} + \sin(a_{other})w_{social} + \sin(a_{memory})w_{memory} + \sin(a_{continuity})w_{continuity}$$

(8) $$\bar{x} = \cos(a_{goal})w_{goal} + \cos(a_{other})w_{social} + \cos(a_{memory})w_{memory} + \cos(a_{continuity})w_{continuity}$$

Software was implemented in Python (version 3.8.10) (Van Rossum & Python Community, 2021) (for tutorials, see (Dalmaijer, 2017; Oliphant, 2007)), using external libraries Matplotlib (version 3.4.3) (Hunter, 2007), NumPy (version 1.21.3) (Harris et al., 2020), SciPy (version 1.7.1) (Virtanen et al., 2020), and utm (version 0.7.0) (Bieniek, 2020).

**Experimental design**

Agents travelled in three conditions that mapped onto work in pigeons (Sasaki & Biro, 2017): solo, paired, and in an experimental condition with generational turnover. In the solo and pair conditions, one or two agents made 60 consecutive journeys. In the experimental condition, a naive replaced an experienced agent every 12 journeys. A total of 10 clean runs were done for each condition, for each unique combination of parameters, resulting in a total of 32370 simulations.

Agents travelled 70 distance units per 1 time unit. While these values were arbitrarily chosen, they impact sampling frequency, and thus parameter estimates. Agents travelling at lower velocities sample the mixture model with less distance in between samples. This should result in a higher continuity weight, to stabilise the route. In sum, while units of both distance and time are arbitrary, locations of peaks in weight parameter space are specific to the current settings.

Weight parameters were varied in a wide and a narrow space. In the wide range, $w_{goal}$ and $w_{social}$ varied from 0.1 to 0.7 in steps of 0.05, and $w_{memory}$ from 0.05 to 0.7 in steps of 0.05, resulting in 557 unique combinations. The narrow range aimed to zoom in on where route efficiency and generational efficiency increase were best. In this narrow range, $w_{goal}$ and $w_{social}$ varied from 0.025 to 0.25 in steps of 0.025, and $w_{memory}$ from 0.025 to 0.7 in steps of 0.025, resulting in 2680 unique combinations.

Spread parameters were fixed at $\kappa_{continuity}$=8.69 (equivalent SD=0.35), $\kappa_{goal}$=1.54 (1.0), $\kappa_{social}$=2.18 (0.80), $\kappa_{memory,1}$=0.85 (2.0) to $\kappa_{memory,5}$=6.78 (0.40), based on model fits for stable pigeon pairs (see under "*Data reduction and statistics*"). This data (Valentini et al., 2021b) was published alongside an analysis on leadership in pairs of naive and experienced pigeons (Valentini et al., 2021a), and seems to have been the source data for an earlier publication on generational improvements in efficiency (Sasaki & Biro, 2017).

**Data reduction and statistics**

Individual pigeon GPS data (defined by latitude and longitude) published by others (Valentini et al., 2021b) was converted to Universal Transverse Mercator (UTM) coordinates (grid zone 30U). Samples with velocities under 25 or over 150 km/h were excluded from flights, to filter breaks and apparent GPS glitches. Flights were completely excluded it they contained coordinates further than 17.03 km (twice the start-goal distance) away from the point midway between start and goal. Finally, flight coordinates were reduced by taking every 20$^{th}$ sample, to fit agent velocity settings (the average inter-sample distance was 3.5 meters, whereas agents' step size was set to 70). No further exclusions were done, and incomplete flights were not imputed. This method subtly deviates from the original paper, but the pattern of results is the same (Figure 2, bottom).

Best parameter fits for pigeon flight data were determined through maximum likelihood estimation. This is an established way of deriving parameter estimates for mixture models of Von Mises distributions, for example in research on visual short-term memory (Bays et al., 2009). For pigeons (N=12) flying in stable pairs, average parameter estimates were $w_{continuity}$=0.58, $w_{goal}$=0.14, $w_{social}$=0.16, and $w_{memory}$=0.12; with $\kappa_{continuity}$=8.14 (equivalent SD=0.36) , $\kappa_{goal}$=1.54 (1.0), $\kappa_{social}$=2.10 (0.82), $\kappa_{memory,1}$=0.28 (1.98) to $\kappa_{memory,5}$=6.83 (0.40).

Simulation results were averaged between paired agents and over independent runs within the same condition and parameter settings. Efficiency for the final generation was computed as the highest out of 12 journeys in the 5$^{th}$ generation. Efficiency and parameters were then averaged for the ten highest final-generation route efficiencies within each condition.

Generational efficiency improvement was computed as the average difference in route efficiency between consecutive generations. To reduce the impact of random fluctuations, the most efficient (typically the final) routes were taken as representative within each generation. The first generation in the experimental condition was omitted, to avoid comparisons between single and paired journeys.

To avoid trivial statistical significance that can be achieved through increasing the number of simulations, inferences on the basis of statistical tests were avoided, and were instead made on the basis of holistic interpretation. Readers are invited to scrutinise figures, data, models, and software.

**Data Availability**

All code and data has been made publicly available through open repositories on GitHub (https://github.com/esdalmaijer/artificial_navigators) and Zenodo (https://doi.org/10.5281/zenodo.6944184).

## Supplementary Figures

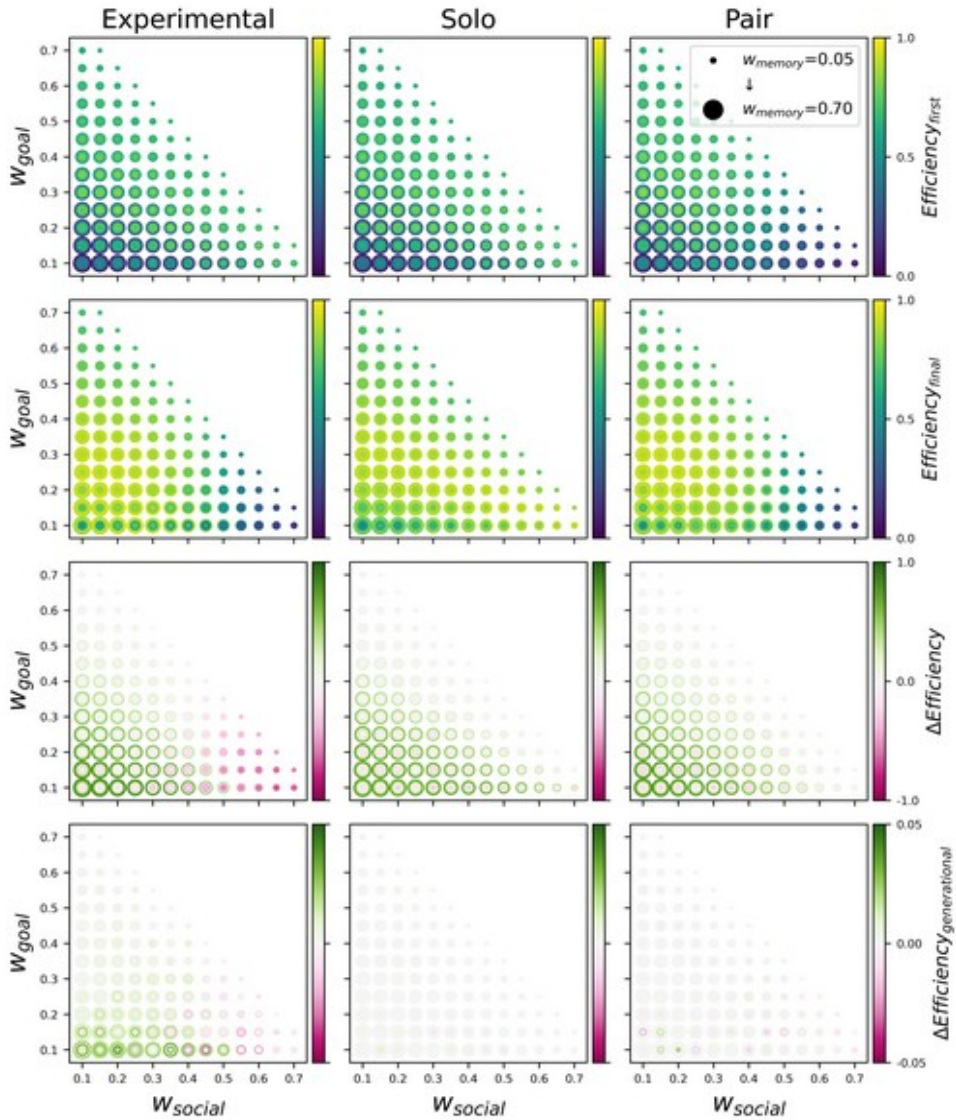

***Figure S1*** *– Each panel shows a measure of route efficiency as a function of $w_{goal}$ (upwards on the y-axis), $w_{social}$ (rightward on the x-axis), and $w_{memory}$ (outwards on the disks). The first row shows the efficiency of agents' first journeys, with lighted colours indicating higher route efficiency. The second row shows the efficiency for agents' final journey (after 5 generations with 12 journeys each), again with lighted colours indicating better route efficiency. The third row shows the increase (green) or decrease (pink) in efficiency between first and final journey. The fourth row shows the average increase (green) or decrease (pink) between consecutive generations. Efficiency was computed as route length divided by Cartesian distance between start and goal. In the experimental condition, a naive agent replaced an experienced one in each generation; in the solo condition, a single agent made all journeys with no generational turnover; and in the pair condition, two agents journeyed together without turnover.*

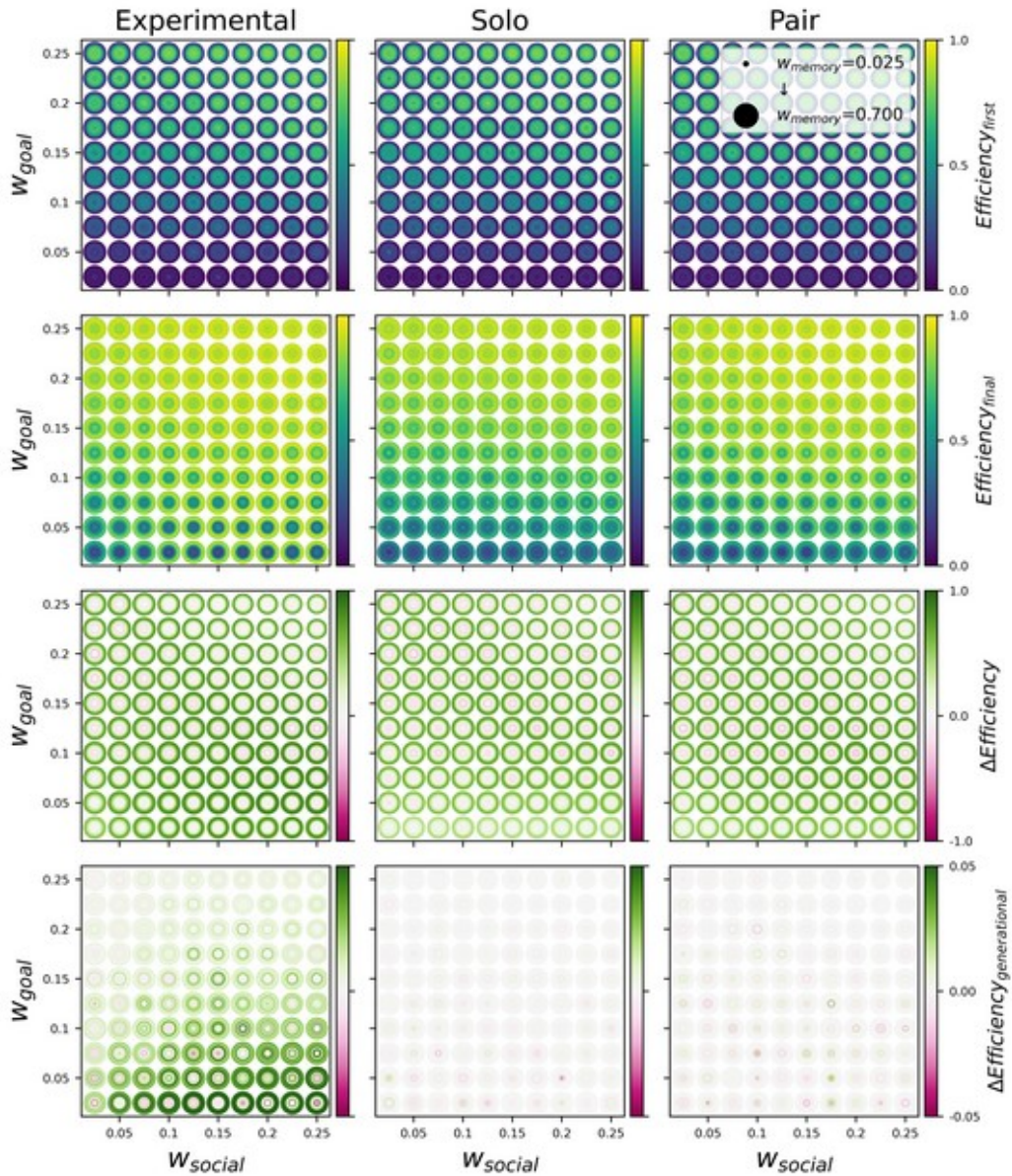

***Figure S2*** *– This figure is similar to Figure S1, but zoomed in on a more narrow parameter space. Each panel shows a measure of route efficiency as a function of $w_{goal}$ (upwards on the y-axis), $w_{social}$ (rightward on the x-axis), and $w_{memory}$ (outwards on the disks). The first row shows the efficiency of agents' first journeys, with lighted colours indicating higher route efficiency. The second row shows the efficiency for agents' final journey (after 5 generations with 12 journeys each), again with lighted colours indicating better route efficiency. The third row shows the increase (green) or decrease (pink) in efficiency between first and final journey. The fourth row shows the average increase (green) or decrease (pink) between consecutive generations. Efficiency was computed as route length divided by Cartesian distance between start and goal. In the experimental condition, a naive agent replaced an experienced one in each generation; in the solo condition, a single agent made all journeys with no generational turnover; and in the pair condition, two agents journeyed together without turnover.*

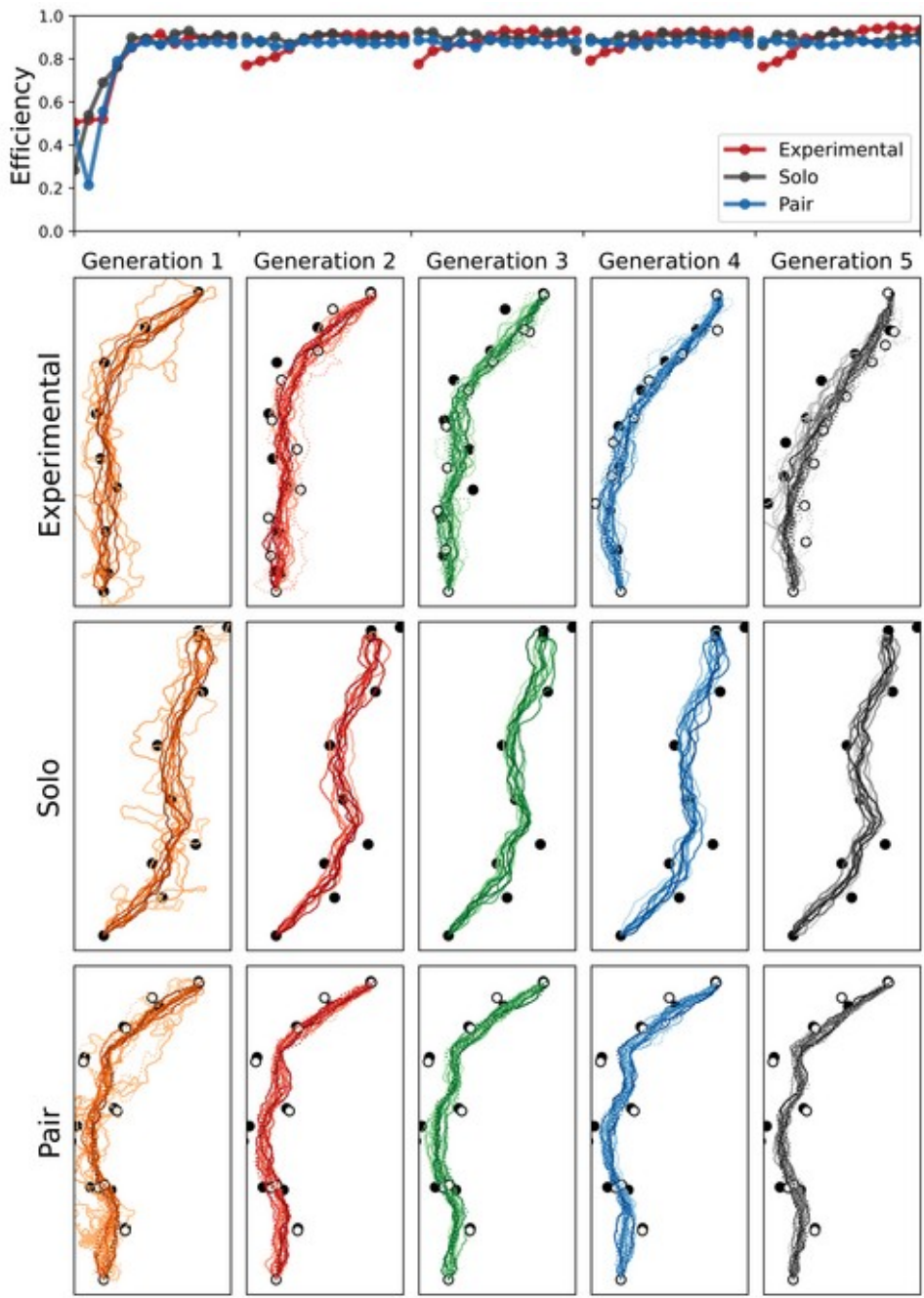

***Figure S3*** *– This figure shows an example of a single complete run through all conditions and generations. The top row shows the route efficiency for each path. The corresponding paths are drawn rows 2-4 (second: experimental condition, third: solo control condition, third: pair control condition). Columns indicate consecutive generations (with experienced-to-naive turnover in the experimental condition, and no turnover in the control conditions). Lighter lines indicate earlier journeys (1-12 within each generation). Solid lines show the first agent, and dotted lines the second. In generations 2-5 in the experimental condition, the first is the experienced agent, and the second is naive. Black dots indicate route landmarks memorised by the first agent, and white dots for the second agent. Crucially, only in the experimental condition, landmarks slowly converge towards the optimal route from start (top right) to goal (bottom left).*

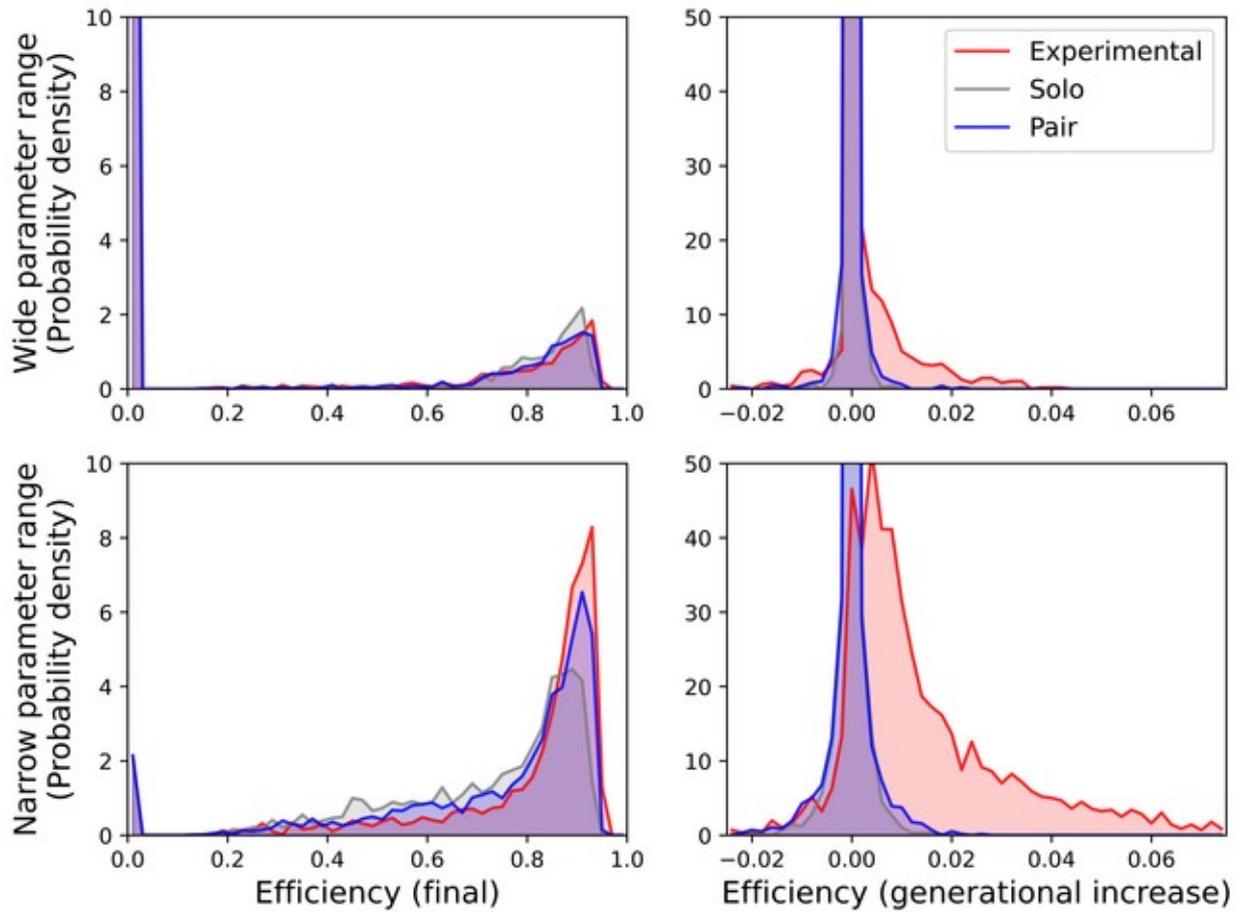

***Figure S4*** *– Histograms of the final-path efficiency (left column) and the mean generational increase in efficiency (right column) for the wide parameter range (top row) and the narrow parameter range (bottom row). In the experimental condition (red), a naive agent replaced an experienced one in each generation; in the solo condition (grey), a single agent made all journeys with no generational turnover; and in the pair condition (blue), two agents journeyed together without turnover. The histograms were computed over all unique combinations of parameters, each represented as the mean over 10 independent runs. For both parameter ranges, the experimental condition shows more combinations of parameters for which final-path efficiency and generational increase in efficiency are relatively high.*

## Version history

**v4, 2023-07-24**

- **Reorganisation of results.** Sub-headings were reintroduced to improve readability. The quantification of cumulative culture and its exclusive occurrence in the experimental condition were stressed to improve clarity. Reported parameters associated with the results displayed in Figure 2 were moved upwards.

- **Addition of control experiments.** In these control experiments, agents with best-result parameters were lesioned to investigate which of the artificial navigation model were essential to cumulative culture.

**v3, 2022-07-30**

- **Reduction of the introduction.** The introduction was condensed to reduce the word count.

- **Removal of sub-headings from Results.** To further reduce word count.

- **Expansion of the Discussion.** To highlight the findings and better embed them in the literature, the Discussion was expanded.

**v2, 2022-06-29**

- **Methodological detail in introduction**. In an attempt to slightly improve the algorithmic detail in the introduction despite a tight word limit, references to the "*Materials and Methods*" section were included.

- **Reorganisation of results**. Previously, figures 2 and 3 related to the optima in parameter space for route efficiency and generational increase in efficiency, respectively. These were moved to the "*Supplementary Information*" (now figures S1 and S2), to prioritise results relating to the increase in efficiency within and between generations. The results are still described in words in the main manuscript, but figures had to be moved due to a limit on display elements in the intended journal.

- **Clarification of parameter space visualisations**. Additional detail was included in the captions of figures S1 and S2, to clarify their colour schemes.

- **How naive agents benefit from experienced navigators**. Under the heading "*Naive agents benefit by copying established routes*" in the "*Results and Discussion*" section, the relative efficiency benefit (compared to the pair condition) for naive individuals is now visualised in Figures 3.

- **How experienced navigators benefit from naive agents**. A new visualisation (Figure 4) was added under the heading "*Experienced agents benefit from regression to the goal*". It shows the distribution of relative bearings towards the naive agent from the perspective of

- the experienced agent. This illustration of regression to the goal in data further supports the theoretical argument already put forward in the previous version.

- **Route accumulation between generations**. The inter-generational shift in routes is now visualised in S3. It shows how landmarks shift in each of the three conditions, due to naive agents copying experienced agents and regression to the goal.

**v1, 2022-06-13**

- Original submission to arXiv.